\documentclass[12pt]{amsart}
\usepackage{amsmath,amsfonts,amssymb} \usepackage{color}
\usepackage[all,cmtip]{xy}
\usepackage{graphicx}
 \usepackage{youngtab}

\newcommand{\corr}[1]{\langle {#1} \rangle}

   \newcommand{\bs}{{\bf s}}
 
\newcommand{\bt}{{\bf t}}
\newcommand{\bT}{{\bf T}}
 \newcommand{\cL}{\mathcal{L}}
\newcommand{\cM}{\mathcal{M}}
  
 \newcommand{\cP}{\mathcal{P}} \newcommand{\cQ}{\mathcal{Q}}
 \newcommand{\cB}{\mathcal{B}}  
  
\newcommand{\rI}{\mathrm{I}} \newcommand{\rII}{\mathrm{II}}
\newcommand{\rIII}{\mathrm{III}} \newcommand{\rIV}{\mathrm{IV}}
\newcommand{\rV}{\mathrm{V}} \newcommand{\rVI}{\mathrm{VI}}
\newcommand{\rVII}{\mathrm{VII}} \newcommand{\rVIII}{\mathrm{VIII}}
  
 \newcommand{\bZ}{\mathbb{Z}}
 \newcommand{\bC}{\mathbb{C}}
 \newcommand{\pd}{\partial}
\newcommand{\Mbar}{\overline{\mathcal M}}

 \DeclareMathOperator{\Ad}{Ad}  \DeclareMathOperator{\ad}{ad}

 \DeclareMathOperator{\Gr}{Gr}
 
\DeclareMathOperator{\ord}{ord}

\newcommand{\be}{\begin{equation}}
\newcommand{\ee}{\end{equation}}
\newcommand{\bea}{\begin{eqnarray}}
\newcommand{\eea}{\end{eqnarray}}
\newcommand{\ben}{\begin{eqnarray*}}
\newcommand{\een}{\end{eqnarray*}}

\newcommand{\half}{\frac{1}{2}}

\newtheorem{cor}{Corollary}[section]

 \newtheorem{thm}[cor]{Theorem}
\theoremstyle{remark}

\definecolor{A}{rgb}{.75,1,.75}

\definecolor{green}{rgb}{0,1,0}
\definecolor{yellow}{rgb}{1,1,0}
\definecolor{orange}{rgb}{1,.7,0}
\definecolor{red}{rgb}{1,0,0}
\definecolor{white}{rgb}{1,1,1}

 \makeindex
\begin{document}
\title
{Emergent Geometry of KP Hierarchy}

\author{Jian Zhou}
\address{Department of Mathematical Sciences\\Tsinghua University\\Beijng, 100084, China}
\email{jzhou@math.tsinghua.edu.cn}

\begin{abstract}
We explain how to construct a quantum deformation of a spectral curve to a tau-function of the KP hierarchy.
This construction is applied to Witten-Kontsevich tau-function  to give a natural explanation
of some earlier work.
We also apply it to higher Weil-Petersson volumes
and Witten's r-spin intersection numbers.
\end{abstract}

\maketitle

\section{Introduction}

By emergent geometry we mean the geometric structures that arise
when one considers the partition function of a Gromov-Witten type theory
on the big phase space.
Such a theory has a finite-dimensional parameter space called the small phase space,
and by the big phase space we mean its formal jet space.
If $\{t_a\}_{a=1}^m$ is a set of local coordinates on the small phase,
one can introduce the induced coordinates $\{t_{a,n}\}_{1 \leq a \leq m, n \geq 0}$
on its formal jet space.
When $n\geq 1$,
$t_{a,n}$ is called a descendant variable.
The partition function  of the theory is a formal power series $Z(\bt)$
in the variables $\{t_{a,n}:\; 1 \leq a \leq m, n \geq 0\}$.
The free energy $F$ is the logarithm of $Z$: $F= \log Z$.
Physically,
one starts with some geometric object and construct an $N=2$ superconformal
field theory from it.
One takes suitable topological twist to get a topological field theory.
The small phase space is the space of coupling constants
of the primary chiral fields,
whereas the big phase space is the space of coupling constants
of their gravitational descendants.
The partition function is generating function of all correlators.
Constructing the mathematical theory and identifying the correlators
as topological objects are often highly technical and often regarded
as significant mathematical achievements.
Mathematically,
one considers the moduli spaces of algebraic curves (with extra structures,
e.g. W-structures or maps into some ambient spaces).
The correlators are the intersection numbers of some cohomology classes on the moduli spaces.

The simplest Gromov-Witten type theory is the Gromov-Witten theory of a point.
Physically it is the theory of topological 2D gravity,
and mathematically it corresponds to the intersection theory
of the Deligne-Mumford moduli spaces of stable algebraic curves.
A famous result, called Witten Conjecture/Kontsevich Theorem \cite{Wit, Kon},
states that the partition function $Z_{WK}$ of topological 2D gravity is a tau-function
of the KdV hierarchy and it satisfies the Virasoro constraints \cite{DVV}.
This result have been generalized to many other interesting cases,
whose partition functions $Z$ are the tau-functions of some integrable hierarchies.
They satisfy Virasoro constraints, or more generally, W-constraints.
More recently,
many of them have been shown to satisfy the topological recursions
starting from some plane curves called the spectral curves \cite{EO}.
One can also use the boson-fermion correspondence to study the partition  functions.
This leads to conformal field theories on the spectral curves \cite{Zho-Emergent}.

In our earlier work on Witten-Kontsevich tau-function,
the following picture for the interrelationships of such results has emerged:
$$
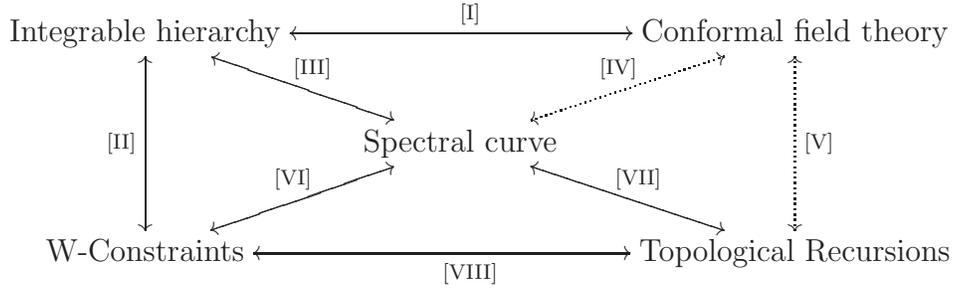
\xymatrix{
\text{Integrable hierarchy} \ar[rr]^{[\rI]}  \ar[dr]^{[\rIII]} \ar[dd]_{[\rII]}
 &  &    \text{Conformal field theory} \ar[ll] \ar@{.>}[dl]_{[\rIV]}   \ar@{.>}[dd]^{[\rV]} \\
                & \text{Spectral curve} \ar[dr] \ar[dl] \ar[ul] \ar@{.>}[ur]             \\
\text{W-Constraints} \ar[ur]^{[\rVI]} \ar[rr]_{[\rVIII]} \ar[uu]
& &    \text{Topological Recursions}  \ar@{.>}[uu] \ar[ul]_{[\rVII]} \ar[ll]     }
$$
Together with some other result which we will briefly review momentarily,
they form what we mean by emergent geometry in the case of Witten-Kontsevich tau-function.
We borrow the terminology from statistical physics where an emergent phenomenon
means a phenomenon that only arises when one has an infinite degree of freedom.

In the center of this picture is some plane curve called the spectral curve.
In the case of Witten-Kontsevich tau-function,
it is the Airy curve defined by:
\be
y = \half x^2.
\ee
It first appeared in the setting of Eynard-Orantin topological recursion
\cite{EO, Eyn, BCSW, Zho-DVV},
establishing [VII] in the above picture.
In the approach of \cite{Zho-DVV},
we note that the EO topological recursions for the Airy curve is equivalent to the DVV Virasoro constraints.
This establishes [VI].
In \cite{Zho-Emergent} we note that the Laurent series:
\be
x = \xi - \frac{t_0}{\xi} - \sum_{n \geq 0} (2n+1)!! \frac{\pd F_0}{\pd t_n}(t_0, 0, \dots) \cdot \xi^{-2n-3},
\ee
where $\xi^2 = 2y$,
leads to the  miniversal deformation of the Airy curve
\be \label{eqn:Def Airy}
y = \half x^2 + t_0.
\ee
Here $t_0$ is a linear coordinate on the small phase space.
Using all possible parameters from the big phase space,
we construct a special deformation of the Airy curve of the following form:
\be \label{eqn:Special def}
x(\xi): = \xi - \sum_{n \geq 0} \frac{t_n}{(2n-1)!!} \xi^{2n-1}
- \sum_{n \geq 0} (2n+1)!!\frac{\pd F_0}{\pd t_n}(\bt) \cdot \xi^{-2n-3}.
\ee
It is uniquely  characterized by the following property,
equivalent to Virasoro constraints in genus zero:
\be
(x(\xi)^2)_- = 0.
\ee
To extend this to arbitrary genera,
we endow the space   of series of the form
\be
\sum_{n =0}^\infty (2n+1) \tilde{u}_n z^{(2n-1)/2}
+ \sum_{n =0}^\infty \tilde{v}_n z^{-(2n+3)/2}
\ee
 the following symplectic structure:
\be
\omega = \sum_{n =0}^\infty   d\tilde{u}_n \wedge d \tilde{v}_n.
\ee
Consider the canonical quantization:
\be
\hat{\tilde{u}}_n = \frac{t_n-\delta_{n,1}}{(2n+1)!!} \cdot, \;\;\;
\hat{\tilde{v}}_n = (2n+1)!! \frac{\pd}{\pd t_n}.
\ee
Corresponding to the field $x$,
we consider the following field of operators on the Airy curve:
\be
\hat{x}(z) = - \sum_{m \in \bZ} \beta_{-(2m+1)} z^{m-1/2}
= - \sum_{m \in \bZ} \beta_{2m+1} z^{-m-3/2}
\ee
where the operators $\beta_{2k+1}$ are defined by:
\be
\beta_{-(2k+1)} = (2k+1) \frac{t_k-\delta_{k,1}}{(2k+1)!!} \cdot, \;\;\;\;
\beta_{2k+1} = (2k+1)!! \frac{\pd}{\pd t_k}.
\ee
A regularized products $\hat{x}(z)^{\odot 2}$ is defined in {\em loc. cit.}, ,
and one can show that the DVV Virasoro constraints  for
Witten-Kontsevich tau-function
is equivalent to the following equation:
\be
(\hat{x}(z)^{\odot 2})_- Z_{WK} = 0.
\ee
In other words,
DVV Virasoro constraints can be rephrased in terms of some quantization of
\eqref{eqn:Special def}.
This establishes [VI] in the direction of obtaining the constraints from the spectral curve.
The reversed direction was also established in {\em loc. cit.}.
where we use the Virasoro constraints in genus zero to compute
the genus zero one-point function restricted to the small phase space.
The relationship [II] was established by \cite{DVV} and \cite{Kac-Sch}.
It states that the $\tau$ is a tau-function of the KdV hierarchy satisfying
the puncture equation iff it satisfies the Virasoro constraints.
In the proof of Dijkgaaf-Verlinde-Verlinde \cite{DVV},
one first derives the string equation from the KdV hierarchy and the puncture equation.
In the proof of Kac-Schwarz \cite{Kac-Sch},
Sato's Grassmannian and boson-fermion correspondence were used.
By considering the fermionic reformulation of the puncture equation,
the point in the Sato grassmannian corresponding to the Witten-Kontsevich
tau-function is characterized using Airy functions,
solution of the Airy equation which is itself a quantization of the Airy curve.
(Airy functions seem to be ubiquitous in Gromov-Witten theory, see \cite{Bur-Jan-Pan} for a survey.
They have appeared in the proof of Witten Conjecture by Kontsevich \cite{Kon},
Okounkov \cite{Oko}, Faber-Zagier relations \cite{Pan-Pix}, Pixton relations \cite{Pan-Pix-Zvo},
and more recently,
in the theory of BCOV holomorphic anomaly equations \cite{Ali-Yau-Zho}. )
Inspired by this work the author has obtained the explicit fermionic formula for $Z_{WK}$ in \cite{Zho-Explicit, Bal-Yan}.
This formula forms the basis of subsequent work \cite{Zho-Emergent}
where a conformal field theory is associated to a tau-function of the KP hierarchy,
hence establishing the relation [I].
In the same work we also explain how to regard the KdV hierarchy as
noncommutative deformation of the Airy curve,
hence establish half of the relation [II] by showing
quantization of the spectral curve leads to the Lax operator in the KdV hierarachy.
The purpose of this work is to complete the development of relation [II]
in the reversed direction.
As in {\em loc. cit.},
we will work with the more general setting of KP hierarchy
and understand how a spectral curve can be associated to a tau-function.
In the mathematics literature,
algebraic geometers have studied in the 1980s the problems of obtaining
the spectral curve of integrable hierarchy,
reversing Krichever's construction of a tau-function from a curve \cite{Kri}.
See \cite{Mul} for a survey.
Our approach to this problem is different.
We will base on the work of Takasaki and Takebe \cite{Tak-Tak}.
These authors consider not only the Lax operator $L$,
but the Orlov-Schulman operator $M$.
They satisfy the string equation:
\be
[L, M] = \hbar,
\ee
which is just a Heisenberg commutation relation.
We understand them as noncommutative Darboux coordinates of a noncommutative symplectic
structure on a noncommmutative plane.
We introduce a $\hat{S}$-operator as a noncommutative generating function
that leads to a system noncommutative Hamilton-Jacobi equations.
Another useful concept introduced by these authors is that of a twistor data
of a tau-function of the KP hierarchy.
It consists of a pair of functions $(f(\hbar, x, \hbar\pd_x)$ and $g(\hbar, x, \hbar \pd_x)$ such that
\be
[f, g] = \hbar,
\ee
and $P = f(\hbar, M, L)$ and $Q = g(\hbar, M, L)$ are differential operators.
We take a noncommutative canonical change of coordinates from $(L, M)$ to $(P, Q)$,
and regard $\hat{S}$ as a generating function in $P, Q$.
We understand the noncommutative Lagrangian submanifold on the noncommutative $(P, Q)$-plane
as the quantum deformation of the special deformation of the spectral curve.
The latter is obtained as follows.
Consider the dispersionless limit of everything (which was the original setting
for the work of {\em loc. cit.}),
one gets an $S$-function on a $(p, q)$-plane with standard symplectic structure.
Consider the Lagrange submanifold defined by this $S$-function.
The $S$-function depends on all parameters in the large phase space.
When restricted to the small phase space,
it gives us the spectral curve.
Hence by turning all parameters on the big phase space one gets
the special deformation.
We will focus on some special twistor data in this paper,
i.e., we require the operator $P$ to be a differnetial operator of finite order.
Then one can use the theory of string equations developed by Schwarz \cite{Sch} based on \cite{Kac-Sch}.

We summarize [III] in the following flow diagram:
$$
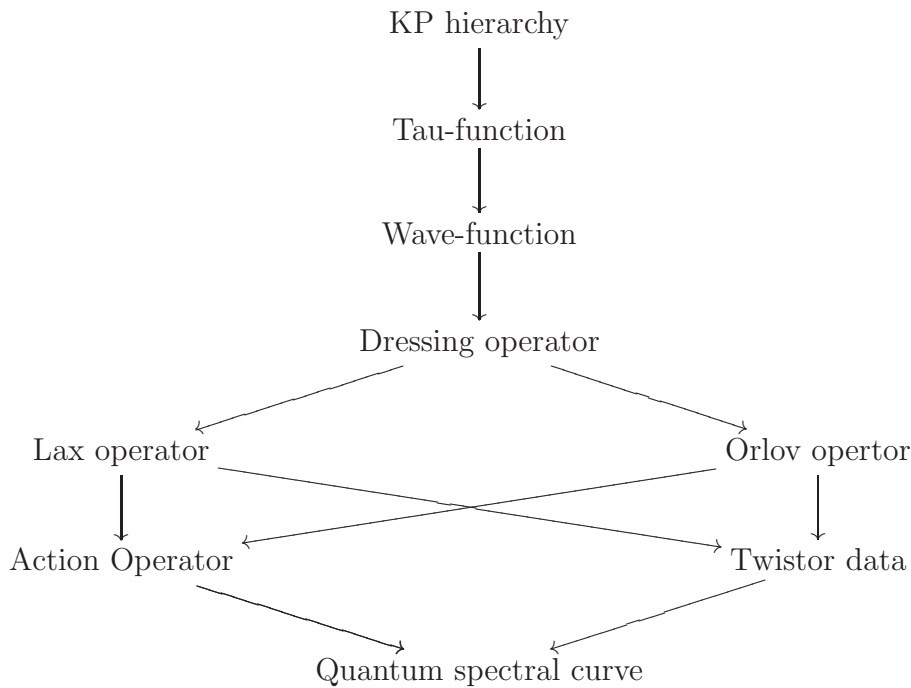
\xymatrix{
& \text{KP hierarchy}  \ar[d] & \\
& \text{Tau-function} \ar[d] & \\
& \text{Wave-function} \ar[d] & \\
& \text{Dressing operator} \ar[ld] \ar[rd] & \\
\text{Lax operator} \ar[d] \ar[rrd] &  &    \text{Orlov opertor} \ar[d] \ar[lld]  \\
\text{Action Operator} \ar[dr] \ar[dr] & & \text{Twistor data} \ar[dl]             \\
&    \text{Quantum spectral curve}    &   }
$$
There is a similar diagram in the dispersionless limit.
In this way,
we not only give a natural explanation of our earlier work on Witten-Kontsevich tau-function,
but also make it very easy to generalize to other situations,
such as higher Weil-Petersson volumes and
Witten's r-spin intersection numbers.
So far we have not been directly established the relationship [IV] and [V].
We hope to address them in future work.

We arrange the rest of the paper as follows.
In Section \ref{sec:KP} we explain the noncommutative symplectic geometry that emerges
from a tau-function of the KP hierarchy.
In Section \ref{sec:String equation} we explain how to combine this with
the theory of string equations.
The diespersionless limits are discussed in Section \ref{sec:Dispersionless}.
As applications,
we discuss the examples of Witten-Kontsevich tau-function, higher Weil-Petersson volumes and
Witten's r-spin intersection numbers in Section \ref{sec:Examples}.
In the final Section \ref{sec:CPS},
we draw some conclusions,
describe some generalizes,
and make some speculations.

\section{Emergent Geometry of KP Hierarchy} \label{sec:KP}

In this section we explain the procedure to construct from a tau-function of
the KP hierarchy a family of quantum spectral curves
on a quantum plane, parameterized by the big phase space.
We are based on Takasaki-Takebe \cite{Tak-Tak} and interpret their results
in the language of quantum symplectic plane,
and use the notion of twistor data introduced by them to construct quantum spectral curves
as quantum plane curves.

\subsection{KP hierarchy}

Suppose that $\tau(\bT; \hbar)$ is a tau-function of the KP hierarchy,
where $\bT = (T_1, T_2, \dots)$,
and $\hbar$ is a formal genus tracking parameter.
The latter is a system of partial differential equations:
\be \label{eqn:KP}
\hbar \frac{\pd}{\pd t_k} L = [B_k, L], \;\;\;\; k = 1, 2, \dots,
\ee
for the pseudo-differential operator $L$ of the form
\be \label{eqn:L}
L = \hbar\pd_x + \sum_{n=1}^\infty u_{n+1}(\bT; \hbar) (\hbar \pd_x)^{-n},
\ee
where $x=T_1$, $\pd_x = \frac{\pd}{\pd T_1}$,
and $B_k$ is a  differential operator defined by
\be \label{eqn:Bk}
B_k = (L^k)_+.
\ee
The coefficients $u_n$ are supposed to be of the following form:
\be
u_n(\bT; \hbar) = \sum_{i=0}^\infty u_n^{(i)}(\bT) \hbar^i.
\ee
This system is equivalent to the Zakharov-Shabat zero-curvature equations:
\be \label{eqn:ZS}
\hbar\frac{\pd B_m}{\pd T_n} - \hbar\frac{\pd B_n}{\pd T_m}
+ [B_m, B_n] = 0.
\ee

\subsection{Wave-function and dressing operator}

By Sato's formula,
one can define the wave function from the tau-function as follows:
\be \label{eqn:Sato}
w(\bT; \xi) = \exp \biggl(\hbar \sum_{n=1}^\infty  T_n \xi^n \biggr) \cdot
\frac{\tau(T_1-\hbar\xi^{-1}, T_2-\hbar\frac{1}{2} \xi^{-2}, \dots; \hbar)}{\tau(T_1, T_2, \dots; \hbar)}.
\ee
It is a formal solution of the form
\be \label{eqn:w}
w = \exp \biggl( \hbar^{-1} \sum_{n=1}^\infty T_n \xi^n \biggr) \cdot
\biggl( 1 + \frac{w_1}{\xi} + \frac{w_2}{\xi^2} + \cdots \biggr)
\ee
to the system
\bea
&& L w= \xi \cdot w,  \label{eqn:L-w} \\
&& \hbar\frac{\pd w}{\pd T_n} = B_n w, \;\; n =1, 2,\dots.  \label{eqn:dw/dtn}
\eea
The  compatibility  condition of this system is the system \eqref{eqn:KP}.

The dressing operator $W$ is defined by:
\be \label{eqn:W}
W:= 1 + \sum_{n=1}^\infty w_j \pd_x^{-j}.
\ee
The dressing operator $W$ and the wave-function $w$ uniquely determine each other:
\be \label{eqn:Wave}
w = W \exp \biggl(\hbar^{-1} \sum_{n=1}^\infty  T_n \xi^n \biggr).
\ee
The operator $L$ and $W$ are related by:
\be \label{eqn:Dressing}
L : = W \circ \hbar \pd_x \circ W^{-1}.
\ee
The evolution of the operator $W$ is governed by the Sato equation:
\be
\frac{\pd}{\pd T_k} W = - (L^k)_- W.
\ee

\subsection{Orlov-Schulman operator}

Taking $\frac{\pd}{\pd \xi}$ on both sides of \eqref{eqn:Wave},
one gets:
\be \label{eqn:dw/dxi}
\hbar\frac{\pd w}{\pd \xi} = M w,
\ee
where $M$ is the  Orlov-Schulman operator defined by
\be
M = W\biggl( \sum_{n=1}^\infty n T_n (\hbar \pd_x)^{n-1} \biggr) W^{-1}.
\ee
It can be written in the following form:
\be \label{eqn:Orlov}
M = \sum_{n=1}^\infty n T_n L^{n-1}
+ \sum_{n=1}^\infty  v_n(\bT; \hbar) L^{-n-1}.
\ee
This operator satisfies the following equations:
\bea
&& \hbar \frac{\pd M}{\pd T_n} = [B_n, M ], \;\; n = 1, 2, \dots, \label{eqn:pd-M-tn} \\
&& [L, M ] = \hbar, \label{eqn:[L,M]}
\eea

The right-hand side of \eqref{eqn:dw/dxi} is
\be
(\sum_{n=1}^\infty n T_n \xi^{n-1}
+ x + \sum_{n=1}^\infty  v_n(\bT; \hbar) \xi^{-n-1}) w;
\ee
on the other,
write
$\tau(\bT; \hbar) = e^{F(\bT; \hbar)}$,
then the left-hand side of \eqref{eqn:dw/dxi} is
\be
\biggl( \sum_{n=1}^\infty n T_n \xi^{n-1}
+ \hbar^2 \sum_{n=1}^\infty \frac{\pd F}{\pd T_n}(\bT-\hbar[\xi^{-1}]; \hbar) \xi^{-n-1}
\biggr) w.
\ee
Hence we obtain the following equation:

\be
\sum_{n=1}^\infty  v_n(\bT; \hbar) \xi^{-n-1}
= \hbar^2 \sum_{n=1}^\infty \frac{\pd F}{\pd T_n}(\bT-\hbar[\xi^{-1}];\hbar) \xi^{-n-1}.
\ee

\subsection{S-function}

Write the logarithm of the wave-function as follows:
\be \label{eqn:log-w}
\log w(\bT; \xi) = \hbar^{-1} \sum_{n=1}^\infty  T_n \xi^n + \hbar^{-1} \sum_{n =0}^\infty S_{n+1}(\bT; \hbar) \xi^{-n}.
\ee
By Sato's formula \eqref{eqn:Sato},
we have
\be \label{eqn:S-n+1}
\hbar^{-1} \sum_{n =0}^\infty S_{n+1}(\bT; \hbar) \xi^{-n}
 = F(\bT-\hbar[\xi^{-1}]; \hbar) - F(\bT; \hbar).
\ee

Consider the inversion of the equation \eqref{eqn:L}:
\be
\hbar \pd_x = L + \sum_{n=1}^\infty q_{n+1}L^{-n}.
\ee
Similarly, one can expand $M$ and $B_n$
in powers of $L$ as follows:
\bea
&& M = \sum_{n=1}^\infty
nT_n L^{n-1} + \sum_{n=1}^\infty v_{n+1} L^{-n-1}, \\
&& B_m = L^m + \sum_{n=1}^\infty q_{m,n+1} L^{-n}.
\eea
Then from the linear equations \eqref{eqn:L-w}, \eqref{eqn:dw/dtn},\eqref{eqn:dw/dxi},
and \eqref{eqn:log-w},
one can deduce the following relations:
\begin{align}
v_{n+1} & = -nS_{n+1}, &
q_{n+1} & = \frac{\pd S_{n+1}}{\pd x}, &
q_{m,n+1} & = \frac{\pd S_{n+1}}{\pd T_m}.
\end{align}
So we have
\bea
&& \hbar \pd = L + \sum_{n=1}^\infty \frac{\pd S_{n+1}}{\pd x} L^{-n},  \label{eqn:d-in-L} \\
&& M = \sum_{n=1}^\infty
nT_n L^{n-1} - \sum_{n=1} n S_{n+1} L^{-n-1},  \label{eqn:M in L} \\
&& B_m = L^m + \sum_{n=1}^\infty \frac{\pd S_{n+1}}{\pd T_m} L^{-n}. \label{eqn:Bm-in-L}
\eea
Note $B_1 = \hbar \pd$,
and so \eqref{eqn:d-in-L} is a special case of \eqref{eqn:Bm-in-L}.

\subsection{Noncommutative Faber polynomials and noncommutative Grunsky coefficients}

Let us briefly recall Faber polynomials and Grunsky coefficients
in the theory of univalent functions.
We will actually use the version
developed in the setting of formal Laurent series by Schur \cite{Schur-Faber}.
Let $g$ be a formal Laurent series in $w$:
$$g(w) =w + \sum_{n=1}^\infty b_n w^{-n},$$
the coefficients $b_n$ are assumed to commute with each other,
and also with $w$.
Consider its Lagrange inversion:
\be \label{eqn:f}
f(z) = z + \sum_{n =1}^\infty a_n z^{-n}.
\ee
See Schur \cite{Schur-Lagrange} for the algebraic version of Lagrange inversion
formula for formal power series.
The $m$-th {\em Faber polynomial} of $f$ or $g$ is
\be
\Phi_m(w) = (g(w)^m)_+,
\ee
where if $g(w)^m = \sum_{i \leq m} b_{m, i} w^i$,
then
\begin{align}
(g(w)^m)_+ & := \sum_{0 \leq i \leq m} b_{m, i} w^i, &
(g(w)^m)_- & := \sum_{i \leq -1} b_{m, i} w^i.
\end{align}
In particular,
\be
\Phi_m(w) =  \sum_{0 \leq i \leq m} b_{m, i} w^i.
\ee
Now note
\ben
&& \Phi_m(f(z)) = g(f(z))^m - \sum_{i < 0} b_{m,i} f(z)^i
= z^m - \sum_{i < 0} b_{m,i} f(z)^i.
\een
So one can write
\be \label{eqn:Grunsky}
\Phi_m(f(z)) = z^m + \sum_{n=1}^\infty c_{mn} z^{-n},
\ee
where $c_{mn}$ are called the {\em Grunsky coefficients}.
Schur \cite{Schur-Faber} derived a beautiful explicit formula for them using algebraic method for Laurent series.
A more compact form of Schur's formula is:
\be
\log \frac{f(z) - f(\xi)}{z -\xi} = -\sum_{m,n=1}^\infty \frac{1}{m} c_{m,n} z^{-m} \xi^{-n}.
\ee
By plugging in \eqref{eqn:f} and expanding the left-hand side as follows:
\ben
&& \log \frac{f(z) - f(\xi)}{z-\xi}
= \log (1 - \sum_{n=1}^\infty a_{n+1}\frac{\xi^{-n+1} - z^{-n+1}}{z-\xi}) \\
& = & - \sum_{K=1}^\infty \frac{1}{K} \biggl(\sum_{n=1}^\infty a_{n+1}\frac{\xi^{-n+1} - z^{-n+1}}{z-\xi}\biggr)^K \\
& = & - \sum_{K=1}^\infty \frac{1}{K} \sum_{2m_2+ \cdots + Km_K=K}
K! \prod_{j=2}^K \frac{a_j^{m_j}}{m_j!}\biggl(\frac{\xi^{-j+1} - z^{-j+1}}{z-\xi}\biggr)^{m_j} \\
& = & - \sum_{K=1}^\infty \frac{1}{K} \sum_{\sum jm_j=K}
K! \prod_{j=2}^K \frac{a_j^{m_j}}{m_j!}\biggl(\frac{\xi/z - (\xi/z)^j}{1-\xi/z}\biggr)^{m_j} \cdot \xi^{-K}.
\een
One can then get
\be \label{eqn:Schur}
\frac{1}{k} c_{k,l}
= (k+1-1)!\sum_{jm_j =k+l} \prod_{j=2}^{k+l} \frac{a_j^{m_j}}{m_j!}
\cdot \biggl[ \prod_{j=2}^{k+l} \biggl( \frac{x-x^j}{1-x}\biggr)^{m_j} \biggr]_{x^k},
\ee
where $[- ]_{x^k}$ denotes the coefficient of $x^k$.
This is the original formula of Schur \cite{Schur-Faber}.

We are now in the natural setting of introducing noncommutative Faber polynomial and Grunsky coefficients.
By \eqref{eqn:L} and \eqref{eqn:Bk},
we understand the operator $B_k$ as the $k$-th noncommutative Faber polynomial of $L$,
it is regarded as polynomial in $\hbar\pd_x$,
with coefficients differential polynomials in $u_{n+1}(x)$, $n=1, 2, \dots$.
We regard \eqref{eqn:d-in-L} as the noncommutative Lagrange inversion of the equation \eqref{eqn:L},
then by comparing \eqref{eqn:Bm-in-L} and \eqref{eqn:Grunsky},
one can regard $q_{m, n+1} = \frac{\pd S_{n+1}}{\pd T_m}$ as the noncommutative Grunsky coefficients.
With such understandings,
we follow some steps in Schur's work \cite{Schur-Faber} in the noncommutative setting.

Let $Q = c_0 (\hbar \pd)^m + c_1 (\hbar \pd)^{m-1} + \cdots + c_m$ be a differential operator of degree $m$,
with coefficients differential polynomials in $u_i$'s.
Then by \eqref{eqn:d-in-L},
one can write:
\be
Q = q_0 L^m + q_1 L^{m-1} + \cdots + q_m + q_{m+1} L^{-1} + \cdots.
\ee
Consider the differential operator
\be
D = Q - ( q_0 B_m + q_1 B_{m-1} + \cdots + q_m).
\ee
By \eqref{eqn:Bm-in-L},
$D$ can be written as series in $L$ with only negative powers,
but since $D$ is a differential operator, it means that $D = 0$.
So one gets:
\be
Q = q_0 B_m + q_1 B_{m-1} + \cdots + q_m.
\ee
Now let
\be \label{eqn:dm-in-L}
(\hbar \pd_x)^m = \sum_{n=0}^\infty a_{m,n} L^{m-n}.
\ee
Then we gets
\be
(\hbar \pd_x)^m = \sum_{n=0}^m a_{m,n} B_{m-n}.
\ee
Now we plug in \eqref{eqn:Bm-in-L} and \eqref{eqn:dm-in-L},
\ben
&& \sum_{n=0}^\infty a_{m,n} L^{m-n}
= \sum_{k=0}^m a_{m,k} (L^k + \sum_{n=1}^\infty q_{k,n+1} L^{-n}).
\een
Comparing the coefficients of $L^{-m}$ on both sides:
\be
a_{m, m+n} = \sum_{k=1}^m a_{m,k} q_{k, n+1}.
\ee
From this one can recursively find $q_{k, n+1}$ in terms of $a_{i,j}$'s,
and ultimately in terms of $q_{n+1} = \frac{\pd S_{n+1}}{\pd x}$.
In other words,
now we know that it is possible to express $q_{k,n+1} = \frac{\pd S_{n+1}}{\pd T_k}$
in terms of $q_{n+1} = \frac{\pd S_{n+1}}{\pd x}$ and their derivatives in $x$.
One might find such expressions by generalizing other steps in \cite{Schur-Faber} to the noncommutative setting
and obtain a noncommutative analogue of Schur's explicit formula.
We will not attempt this here.

\subsection{The $\hat{S}$-operator and noncommutative Hamilton-Jacobi equation}

Introduce an operator
\be \label{def:S-Hat}
\hat{S} := \sum_{n=1}^\infty  T_n L^n
+ \sum_{n =0}^\infty S_{n+1}(\bT; \hbar) L^{-n},
\ee
then one has
\be
\hbar \hat{S} w = (\sum_{n=1}^\infty  T_n \xi^n
+ \sum_{n =0}^\infty S_{n+1}(\bT; \hbar) \xi^{-n}) w.
\ee
One can regard $\hat{S}$ as the {\em noncommutative action operator}.
Treating $\hat{S}$ as a function in independent variables $\bT$ and $L$,
where $T_1 =x$ and $L$ are noncommutative to each other.
Define noncommutative derivatives $\pd_{T_n}$ and $\pd_L$,
such that
\begin{align}
\pd_{T_m} T_n & = \delta_{mn}, & \pd_{T_m} L & = 0, \\
\pd_L T_n & = 0, & \pd_L L & =1.
\end{align}
Then one has
\begin{align} \label{eqn:Noncomm-H-J}
\pd_L \hat{S} & = M, & \pd_{T_n} \hat{S} & = B_n.
\end{align}
Here $L$ is the noncommutative generalized position,
and $M$ is the noncommutative conjugate momentum;
$T_n$'s are time variables, and $B_n$ are the corresponding Hamiltonians.

\subsection{Noncommutative symplectic geometry}

Rewrite \eqref{eqn:Noncomm-H-J} as an equation of the differential form:
\be
d\hat{S} = M dL + \sum_{n=1}^\infty B_n d T_n.
\ee
Take exterior differential one more time:
\be
0 = d^2 \hat{S} = dM \wedge dL + \sum_{n=1}^\infty dB_n \wedge dT_n,
\ee
or equivalently,
\be
-dM \wedge dL = \sum_{n=1}^\infty  dB_n \wedge dT_n.
\ee
Consider the noncommutative 2-form
\be
\omega: = \sum_{n=1}^\infty  dB_n \wedge dT_n.
\ee
Then one has
\be
d \omega = 0
\ee
and
\be
\omega \wedge \omega =0.
\ee
The latter equation is equivalent to the zero-curvature condition \eqref{eqn:ZS}.
So one can think of the pair $(L, M)$ as noncommutative Darboux coordinates for the noncommutative
two-form $\omega$.

\subsection{A grading} \label{sec:Grading}

Following Takasaki-Takebe \cite{Tak-Tak},
define the ¡°order¡± and the ¡°principal symbol ¡± of pseudo-differential operators as follows:
For $A = \sum a_{n, m}(\bT) \hbar^n \pd^m$,
define
\be
\ord^\hbar(A) := \max \{m-n |\; a_{n,m}(\bT) \neq 0\}.
\ee
In particular,
$\ord^\hbar (\hbar) = -1$,
$\ord^\hbar(\pd_x) = 1$.
The {\em principal symbol} and the {\em symbol of order $l$} of a
pseudo-differential operator $A$ are defined by
\bea
&& \sigma^\hbar(A):= \hbar^{-\ord^\hbar(A)} \sum_{m-n= \ord^\hbar(A)}
a_{n,m}(\bT) k^m, \\
&& \sigma^\hbar_l(A): =
\hbar^{-l}
\sum_{m-n= \ord^\hbar(A)}
a_{n,m}(\bT) k^m,
\eea

\subsection{Riemann-Hilbert problem and twistor data} \label{sec:Twistor}

Takasaki and Takebe \cite{Tak-Tak} proved the following  important results.

\begin{thm} (\cite[Proposition 1.7.11, Proposition 1.7.12]{Tak-Tak})
Suppose that
\ben
&& f(\hbar, x, \hbar \pd_x) = \sum_{n\in\bZ}  f_n(\hbar, x)(\hbar \pd_x )^n, \\
&& g(\hbar, x, \hbar \pd_x) = \sum_{n \in \bZ} g_n(\hbar, x)(\hbar \pd_x )^n
\een
are two pseudo-differential operators of $0$-th order,
i.e., $\ord^\hbar f = \ord^\hbar g = 0$,
and that they satisfy the canonical commutation relation $[f, g] = \hbar$.
Assume that there are two pseudo-differential operators $L$   and $M$ of the forms
\ben
&& L = \hbar \pd_x +
\sum_{n=1}^\infty
u_{n+1}(\hbar, t)(\hbar \pd_x )^{-n}, \\
&& M = \sum_{n=1}^\infty
nt_nL^{n-1} + \sum_{n=1}^\infty v_{n}(\hbar , t) L^{-n-1},
\een
respectively,
are given and that
$\ord^\hbar L = \ord^\hbar M = 0$,
$[L, M ] = \hbar$.
Then, if $f(\hbar, M, L)$ and $g(\hbar, M, L)$ are both differential operators,
i.e.,
$$(f (\hbar, M, L))_-
= (g(\hbar, M, L))_-
= 0,
$$
$L$ is a solution of the KP hierarchy, and $M$ is the corresponding Orlov-Shulman operator.
The pair $(f, g)$ above is called the {\em twistor data} of the tau-function
of the KP hierarchy.
Conversely, any solution of the KP hierarchy possesses a twistor data.
\end{thm}

Write $P = f(M, L;\hbar)$ and $Q = g(M, L; \hbar)$.
Then one has
\be
[P, Q] = \hbar.
\ee
Hence one can also use the pair $(P, Q)$ as noncommutative Darboux coordinates.
In particular,
one can express $\hat{S}$ in terms of $P$ and $Q$.

\subsection{Noncommutative Landau-Ginzburg superpotential}

Now the operator $P$ is a differential operator:
\be
P = \sum_{j=0}^m a_j(\bT; \hbar)\cdot (\hbar \pd_x)^j
\ee
for some functions $a_k(\bT; \hbar)$.
We will refer to $P$ as the {\em noncommutative Landau-Ginzburg superpotential}.

\subsection{Noncommutative special deformation}

Let us rewrite $\hat{S}$ as a noncommutative function on the noncommutative $(P, Q)$-plane.
We will consider the following noncommutative curve:
\be
Q = \pd_P \hat{S}.
\ee
We will call this the {\em noncommutative special deformation}.

\section{String Equation and W-Constraints}

\label{sec:String equation}

The operators $P$ and $Q$ coming from the twistor data are differential operators, maybe of infinite orders.
In this Section we will focus on the case of $P$ being a differential operator.
We will review the work of Schwarz \cite{Sch}
which generalizes Kac-Schwarz \cite{Kac-Sch}.

\subsection{Special twistor data and string equation}

Suppose that for some twistor data $P$ is a differential operator of order $n$,
of the following form:
\be
P = (\hbar \pd_x)^n + \sum_{i=0}^{n-2} a_i(x, \hbar) \cdot (\hbar \pd_x)^i.
\ee
Then one can rewrite it in terms of the Lax operator $L$:
\be
P = L^n + \sum_{j \leq n-2} b_j(x, \hbar) \cdot L^j.
\ee
Suppose furthermore that $P = L^n$.
Then by \eqref{eqn:Dressing} one has:
\be \label{eqn:P Dressing}
P = W (\hbar\pd_x)^n W^{-1}.
\ee
By a well-known computation,
\be \label{eqn:Higher Heisenberg}
(\hbar \pd_x)^n x = n \hbar (\hbar \pd_x)^{n-1} + x(\hbar \pd_x)^n.
\ee
If follows that
\be
[(\hbar \pd_x)^n, x (\hbar \pd_x)^{1-n}] = n \hbar.
\ee
So one has:
\be
Q = \frac{1}{n} W
\biggl(x (\hbar \pd_x)^{1-n} + \sum_m c_m(x, \hbar) \cdot (\hbar \pd_x)^m \biggr)
W^{-1}.
\ee
Suppose that there is some $N$ such that $c_m = 0$ for $m > N$.
Then from
\be
[(\hbar \pd_x)^n, \sum_{m \geq N} c_m(x, \hbar)\cdot (\hbar \pd_x)^m ] = 0,
\ee
one can inductively show that $c_m$ are independent of $x$ for all $m$,
hence we will write them simply as $c_m(\hbar)$.
And so
\be \label{eqn:Q Dressing}
Q =  W
\biggl(\frac{1}{n} x (\hbar \pd_x)^{1-n} + \sum_m c_m(\hbar) \cdot (\hbar \pd_x)^m \biggr)
W^{-1}.
\ee
We will call a twistor data with differential operators
$P$ and $Q$ given by \eqref{eqn:P Dressing} and \eqref{eqn:Q Dressing}
special twistor data.
In this case the equation $[P, Q]= \hbar$ is called the  string equation.

\subsection{Sato Grassmannian}

In last subsection we are naturally led to the space  $\bC((\pd_x^{-1}))$
of pseudo-differential operators with constant coefficients.
Now it is a perfect place to recall  Sato Grassmannian.
Denote by $E$ the ring of pseudo-differential operators.
Then $H = E/Ex$ is isomorphic as a vector space to the set $\bC((\pd_x^{-1}))$.
After a Fourier transform,
\be
H \cong \bC((z^{-1})),
\ee
with the action of $x^m \pd_x^n$ transformed to the action of $\pd_z^m z^n$.
Indeed,
this follows from \eqref{eqn:Higher Heisenberg}:
\be
x^m \pd_x^n \cdot \pd_x^l
= (-1)^m m! \binom{n+l}{m}  \pd_x^{n+l-m} \pmod{E x}.
\ee
Let $H_+ = \bC[z]$ and $H_- = z^{-1} \bC[[z^{-1}]]$,
One has a decomposition $H = H_+ \oplus H_-$.
Let $\pi: H \to H_+$ be the projection.
The big cell $\Gr^{(0)}$ of Sato grassmannian
consists of linear subspaces $V$ of $H$ such that
$\pi_+|_V: V \to H_+$ are isomorphisms.

\subsection{Kac-Schwarz operator}

One can use the dressing operator to
associate an element $V$ corresponding to the tau-function as follows:
\be
V = S^{-1} H_+.
\ee
Since $P$ and $Q$ are differential operators,
they satisfy:
\begin{align}
P H_+ & \subset H_+, & QH_+ & \subset H_+.
\end{align}
By \eqref{eqn:P Dressing} and \eqref{eqn:Q Dressing},
these conditions are equivalent to:
\begin{align} \label{eqn:KS}
z^n V & \subset V, & AV \subset V,
\end{align}
where $A$ is the Kac-Schwarz operator:
\be
A = \hbar^{1-n} \frac{\pd}{\pd z^n} + \sum_m c_m(\hbar) \cdot (\hbar z)^m.
\ee

\subsection{W-constraints}

By \eqref{eqn:KS},
one gets:
\be \label{eqn:V}
z^{kn} A^lV \subset V
\ee
for all $k \geq 0$, $l \geq 0$.
One can convert these into constraints on the tau-function.
When $l =0$,
$z^{kn} V \subset V$ for $k \geq 1$ correspond to the condition
\be
\frac{\pd}{\pd T_{kn}} \tau = 0, \;\; k \geq 1.
\ee
When $l = 1$,
$z^{kn} A V \subset V$ correspond to the Virasoro constraints:
\be
L_k = \frac{2n} \sum_{j=1}^{kn-1} \frac{\pd^2}{\pd T_j \pd T_{kn-j}}
+ \frac{n^2-1}{24n} \delta_{k, 0}
+ \sum_m \alpha_m(\hbar) \frac{\pd}{\pd T_{(k+1)n+m}}.
\ee
When $l > 1$,
one gets more general $W$-constraints.

\section{Emergent Geometry of Dispersionless KP Hierarchy}
\label{sec:Dispersionless}

In this Section we recall some facts about the dispersionless
limit of the KP hierarchy.
It turns out that except for the wave-function,
almost everything in Section \ref{sec:KP} has a dispersionless version.
Much of this Section is again based on the survey of Takasaki-Takebe \cite{Tak-Tak}.

\subsection{Poisson structure induced by the grading}

It is easy to see that if $A_i$ is a pseudodifferential operator and $\ord^\hbar(A_i) = a_i$,
$i=1,2$,
then
$\ord^\hbar([A_1, A_2]) \leq a_1+a_2-1$,
and
\be
\sigma^\hbar_{a_1+a_2-1}([A_1, A_2])
= \hbar \{\sigma^\hbar_{a_1}(A_1), \sigma^\hbar_{a_2}(A_2)\}.
\ee
Here the Poisson bracket $\{\cdot, \cdot\}$ is defined by:
\be
\{ \varphi(x, k), \psi(x, k)\}
= \frac{\pd \varphi}{\pd k} \frac{\pd \psi}{\pd x} - \frac{\pd A}{\pd x} \frac{\pd \psi}{\pd k}.
\ee

\subsection{Dispersionless KP hierarchy}

Write
$L = L^{(0)} + L^{(-1)} \hbar + \cdots$,
and $B_k = B_k^{(0)} + B_k^{(-1)} \hbar+ \cdots$,
where $\ord^\hbar L^{(-i)}= \ord^\hbar B_k^{(i)} = 0$, $i\geq 0$,
$k \geq 1$.
By \eqref{eqn:KP},
\be
\frac{\pd}{\pd T_n} L^{(k)} = \sum_{i=0}^k [B_n^{(i)}, L^{(k-i)}].
\ee
In particular,
\be
\frac{\pd}{\pd T_n} L^{(0)} =  [B_n^{(0)}, L^{(0)}].
\ee
Recall
$L^{(0)} = \hbar\pd + \sum_{n=1}^\infty u_{n+1}^{(0)} (\bT) (\hbar \pd)^{-n}$.
Write $$B_n^{(0)} = (\hbar \pd_x)^n + \sum_{i=0}^{n-2} B_n^{(i)}(\bT) (\hbar \pd_x)^i.$$
Associate formal power series $\cL$ and $\cB_n$ to these operators:
\bea
&& \cL = k + \sum_{n=1}^\infty u_{n+1}^{(0)} (\bT) k^{-n},
\label{eqn:L0} \\
&& \cB_n = k^n + \sum_{i=0}^{n-2} B_n^{(i)}(\bT) k^i. \label{eqn:Bn0}
\eea
These are just the principal symbol of the corresponding operators.
One can check that
\be
\cB_n = (\cL^n)_+,
\ee
and by \eqref{eqn:KP},
\be
\frac{\pd \cL}{\pd T_n} = \{\cB_n, \cL\}, \;\; n =1, 2, \dots.
\ee
This is called the dispersionless KP hierarchy.
This system is equivalent to the zero-curvature equations:
\be \label{eqn:ZS0}
\frac{\pd \cB_m}{\pd T_n} - \frac{\pd \cB_n}{\pd T_m}
+ \{\cB_m, \cB_n\} = 0.
\ee
Consider a two-form
\be
\omega := \sum_{n=1}^\infty d \cB_n \wedge dT_n.
\ee
Then one has
\be
d\omega =0,
\ee
and \eqref{eqn:ZS0} is equivalent to
\be
\omega\wedge \omega = 0.
\ee

\subsection{Dispersionless  dressing operator}

Take logarithm on both sides of \eqref{eqn:W}:
\be
\log W:=  \hbar^{-1} X,
\ee
where $X=\sum_{n=1}^\infty \chi_j(\bT; \hbar) (\hbar \pd_x)^{-j}$
and $\ord^\hbar X = 0$ (cf. \cite[Prop. 1.7.5]{Tak-Tak}).
Let $\varphi(\bT) = \sigma^\hbar(X)$,
then by \cite[Corollary 1.7.6]{Tak-Tak},
\be
\cL = e^{\Ad_\varphi}(k),
\ee
where $\Ad_\varphi$ is defined as taking Poisson bracket with $\varphi$:
\be
\Ad_\varphi f(k,x) = \{\varphi, f(k, x)\}
= \frac{\pd \varphi}{\pd k} \frac{\pd f(k,x)}{\pd x}
- \frac{\pd \varphi}{\pd x} \frac{\pd f(k,x)}{\pd k};
\ee
furthermore,
\be
\nabla_{T_n, \varphi} \varphi = -(e^{Ad_\varphi}(k^n))_-, \;\; n =1, 2, \dots,
\ee
where the left-hand side is defined by:
\be
\nabla_{T_n, \varphi} \psi = \sum_{m=0}^\infty
\frac{1}{(m+1)!} (\Ad_\varphi)^m (\frac{\pd \psi}{\pd T_n}) .
\ee

\subsection{Orlov-Schulman function}

The Orlov-Schulman function $\cM$
is the principal symbol of   Orlov-Schulman operator:
\be
\cM = \sigma^\hbar(M).
\ee
It is given by \cite[(1.3.1)]{Tak-Tak}
\be
\begin{split}
\cM = & e^{\ad_\varphi}(\sum_{n=1}^\infty nT_n k^{n-1}) \\
= & e^{\ad_\varphi} e^{\ad t(k)}(x),
\end{split}
\ee
where $t(k) = \sum_{n=1}^\infty n T_n k^n$, $x = T_1$.

The dispersionless version of the equations \eqref{eqn:pd-M-tn} and \eqref{eqn:[L,M]} are
\bea
&& \frac{\pd \cM}{\pd T_n} = [\cB_n, \cM ], \;\; n = 1, 2, \dots,
\label{eqn:pd-M-tn=0} \\
&& [\cL, \cM ] = 1, \label{eqn:[L,M]-0}
\eea
respectively.

Write $F = \sum_{g \geq 0} \hbar^{2g-2} F_g$,
then the dispersionless version of \eqref{eqn:Orlov} is
\be \label{eqn:Orlov-0}
\cM = \sum_{n=1}^\infty n T_n \cL^{n-1}
+ \sum_{n=1}^\infty  v_n^{(0)}\emph{}(\bT) \cL^{-n-1},
\ee
where
\be
\sum_{n=1}^\infty  v_n^{(0)}(\bT) \xi^{-n-1}
=  \sum_{n=1}^\infty \frac{\pd F_0}{\pd T_n}(\bT) \xi^{-n-1}.
\ee

\subsection{Dispersionless S-function}

The genus zero part of $\log w$ will be be called the dispersionless $S$-function
and will be denoted by $S^{(0)}$.
By \eqref{eqn:log-w} and \eqref{eqn:S-n+1},
\be
S^{(0)}(\bT, \xi) = \sum_{n=1}^\infty  T_n \xi^n
+ \sum_{n =0}^\infty S_{n+1}^{(0)}(\bT) \xi^{-n},
\ee
where the coefficients $S_{n+1}^{(0)}$ are given by:
\be
S_{n+1}^{(0)}(\bT) = - \frac{1}{n} \frac{\pd F_0(\bT)}{\pd T_n}.
\ee

Consider the inversion of the equation \eqref{eqn:L0}:
\be
k = \cL + \sum_{n=1}^\infty q_{n+1}^{(0)} \cL^{-n}
\ee
Similarly, one can expand $\cM$ and $\cB_n$
in powers of $\cL$ as follows:
\bea
&& \cM = \sum_{n=1}^\infty
nT_n \cL^{n-1} + \sum_{n=1}^\infty v^{(0)}_{n+1} \cL^{-n-1}, \\
&& \cB_m = \cL^m + \sum_{n=1}^\infty q^{(0)}_{m,n+1} \cL^{-n}.
\eea
Then from the linear equations \eqref{eqn:L-w}, \eqref{eqn:dw/dtn},\eqref{eqn:dw/dxi},
and \eqref{eqn:log-w},
one can deduce the following relations:
\begin{align}
v^{(0)}_{n+1} & = -nS^{(0)}_{n+1}, &
q^{(0)}_{n+1} & = \frac{\pd S^{(0)}_{n+1}}{\pd x}, &
q^{(0)}_{m,n+1} & = \frac{\pd S^{(0)}_{n+1}}{\pd T_m}.
\end{align}
So we have
\bea
&& k = \cL - \sum_{n=1}^\infty \frac{1}{n} \frac{\pd^2 F_{0}(\bT) }{\pd x \pd T_n} \cL^{-n},
\label{eqn:d-in-L0} \\
&& \cM = \sum_{n=1}^\infty
nT_n \cL^{n-1} + \sum_{n=1} \frac{\pd F_0(\bT)}{\pd T_n}  \cL^{-n-1},  \label{eqn:M in L0} \\
&& \cB_m = \cL^m - \sum_{n=1}^\infty \frac{1}{n} \frac{\pd^2 F_0(\bT)}{\pd T_m \pd T_n} \cL^{-n}.
\label{eqn:Bm-in-L0}
\eea

By \eqref{eqn:L0} and \eqref{eqn:Bn0},
$\cB_n$ is the $n$-th Faber polynomial of $L$.
By comparing \eqref{eqn:Bm-in-L0} and \eqref{eqn:Grunsky},
$q^{(0)}_{m, n+1} = \frac{\pd S^{(0)}_{n+1}}{\pd T_m}$ are the Grunsky coefficients.
Such observations was first made by Teo \cite{Teo}.
 One may use Schur's explicit formula \eqref{eqn:Schur} to carry explicit calculations.

\subsection{Hamilton-Jacobi equation}

Now by \cite[Proposition 1.7.10]{Tak-Tak},
under the Legendre transform $(t, \xi) \to (t, k)$ given by
\be
k = \frac{\pd S^{(0)}(\bT, \xi)}{\pd x},
\ee
the spectral parameter $\xi$ can be identified with $\cL$:
\be
\xi = \cL(\bT; k) = k + \sum_{n=1}^\infty u_{n+1}^{(0)} (\bT) k^{-n},
\ee
and $S^{(0)}(\bT; \xi)$ becomes a function in $(\bT, k)$:
\be
S^{(0)}(\bT; k) = \sum_{n=1}^\infty  T_n \cL^n
+ \sum_{n =0}^\infty S^{(0)}_{n+1}(\bT) \cL^{-n}.
\ee
Using this equation,
one can regard $S^{(0)}$ as a function in independent variables $\bT$ and $\cL$.
Then one has
\begin{align} \label{eqn:H-J}
\frac{\pd S^{(0)}}{\pd \cL} & = \cM, &
\frac{\pd S^{(0)}}{\pd T_n} & = \cB_n.
\end{align}

\subsection{Symplectic geometry}

Rewrite \eqref{eqn:H-J} as an equation of the differential form:
\be
d S^{(0)} = \cM d\cL + \sum_{n=1}^\infty \cB_n d T_n.
\ee
Take exterior differential one more time:
\be
0 = d^2 S^{(0)} = d\cM \wedge d\cL + \sum_{n=1}^\infty d\cB_n \wedge dT_n,
\ee
or equivalently,
\be
-d \cM \wedge d\cL = \sum_{n=1}^\infty  d\cB_n \wedge dT_n.
\ee
Consider the exterior 2-form
\be
\omega^{(0)}: = \sum_{n=1}^\infty  d\cB_n \wedge dT_n.
\ee
Then one has
\be
d \omega^{(0)} = 0
\ee
and
\be
\omega^{(0)} \wedge \omega^{(0)} =0.
\ee
The latter equation is equivalent to the zero-curvature condition \eqref{eqn:ZS0}.
So one can think of the pair $(\cL, \cM)$ as Darboux coordinates for the noncommutative
two-form $\omega^{(0)}$.

\subsection{Dispersionless twistor data}

Let two  pseudodifferential operators  $f(x, \hbar \pd_x; \hbar)$
and $g(x, \hbar \pd; \hbar)$
be the twistor data as defined in \S \ref{sec:Twistor}.
Let $f^{(0)}(x,k) = \sigma^\hbar(f)$ and $g^{(0)}(x,k) = \sigma^\hbar(g)$
be their principal symbols respectively.
They satisfy the canonical commutation relation
\be \label{eqn:f-g0}
\{f^{(0)}, g^{(0)}\} = 1
\ee
and furthermore,
\bea
&& (f^{(0)} (\cM, \cL))_- = (g^{(0)}(\cM, \cL))_- = 0.
\eea
The pair $(f^{(0)}, g^{(0)})$ is called the {\em dispersionless twistor data} of the tau-function
of the KP hierarchy.

Write $\cP = f^{(0)}(\cM, \cL)$ and $\cQ = g^{(0)}(\cM, \cL)$.
They satisfy  the following equation:
\be
[\cP, \cQ] = 1.
\ee
Hence one can also use the pair $(\cP, \cQ)$ as Darboux coordinates.
In particular,
one can express $S^{(0)}$ in terms of $\cP$ and $\cQ$.

\subsection{Landau-Ginzburg superpotential and special deformation}

Now  $\cP$ is a polynomial in $k$:
\be
\cP = \sum_{j=0}^m a_j^{(0)}(\bT) k^j
\ee
for some functions $a_j{(0)}(\bT)$.
We will refer to $\cP$ as the {\em Landau-Ginzburg superpotential}
of the tau-function of the KP hierarchy.

Let us rewrite $S^{(0)}$ as a function on the $(\cP, \cQ)$-plane,
with $\bT$ as parameters.
We will consider the following  curve:
\be
\cQ = \pd_{\cP} S^{(0)}.
\ee
We will call this the {\em special deformation of the spectral curve}.

\subsection{Relationship with Section \ref{sec:KP}}

In this section we have taken the point of view of taking semiclassical limits
of everything in last section.
This has been  very successful with the help of taking principal symbols.
It is very interesting to reverse the direction and consider
the procedure of quantizing the structures of this section and recover
the structures in Section \ref{sec:KP}.

The basic geometric setting of this section is the symplectic plane with coordinates
$(k, x)$ and symplectic structure $dk \wedge dx$,
and the algebra of functions on this plane is taken to be the space of Laurent series
\be
\sum_{n \geq n_0} a_n(x) k^n,
\ee
endowed with the Poisson bracket of the above symplectic structure
\be
\{f(k,x), g(k, x)\} = \frac{\pd f}{\pd k} \frac{\pd g}{\pd x}
 - \frac{\pd f}{\pd x} \frac{\pd g}{\pd k}.
\ee
After the Moyal quantization,
the algebra of functions can be identified with
the noncommutive algebra of pseudodifferential operators
\be
\sum_{n \geq n_0} a_n(x) (\hbar \pd_x)^n,
\ee
and the Moyal bracket can be identified with the bracket of pseudodifferential operators.
One is naturally led to a version of noncommutative symplectic geometry as in Section \ref{sec:KP}.

\section{From 2D Topological Gravity to Quantum Mechanics}
\label{sec:Examples}

In this Section we explain how to understand
some earlier papers \cite{Zho-Quant-Def, Zho-Emergent}
from the point of view of this paper.
We also discuss the relationship to the generating series of higher Weil-Petersson volumes
and some generalizations to Witten's r-spin intersection numbers.

\subsection{Partition function of topological 2D gravity}

As pointed by WItten \cite{Wit},
the partition function of topological 2D gravity can be identified with the generating series
of some intersection numbers on  $\Mbar_{g,n}$,
the Deligne-Mumford spaces  of stable algebraic curves of genus $g$,
with $n$ marked points.
Let $\psi_1, \dots, \psi_n$ be the first Chern classes of the cotangent line bundles
corresponding to the marked points.
The correlators of the 2D topological gravity are defined as the following intersection numbers:
\be
\corr{\tau_{a_1}, \cdots, \tau_{a_n}}_g :=
\int_{\Mbar_{g,n}} \psi_1^{a_1} \cdots \psi_n^{a_n}.
\ee
The free energy of 2D topological gravity is defined by
\be
F(\bT; \lambda) = \sum_{g \geq 0} \lambda^{2g-2} F_g(\bT).
\ee
where the genus $g$ part of the free energy is defined by:
\be
F_g(\bT) =
= \sum_{n \geq 0} \frac{1}{n!} \sum_{a_1, \dots, a_n \geq 0} T_{a_1} \cdots T_{a_n}
\corr{ \tau_{a_1}, \cdots, \tau_{a_n}}_g.
\ee
The partition function of 2D topological gravity,
often referred to as the  Witten-Kontsevich tau-function, is defined by
\be
Z_{WK}(\bt, \lambda) = \exp F(\bt, \lambda).
\ee

\subsection{Witten Conjecture/Kontsevich Theorem}

According to Witten \cite{Wit} and Kontsevich \cite{Kon},
$Z_{WK}$ is a tau-function of the KdV hierarchy:
\be \label{eqn:KdV}
\pd_{t_n} u = \pd_{t_0} R_{n+1},
\ee
where $t_0 = x$, $u = \lambda^2 \frac{\pd F}{\pd^2 t_0^2}$,
and $R_n[u]$ is a sequence of differential polynomials in $u$ defined recursively by:
\be
\begin{split}
& R_1 = u, \\
& \frac{\pd}{\pd x} R_{n+1} = \frac{1}{2n+1} \biggl(\pd_xu \cdot R_n
+ 2u \cdot \pd_x R_n + \frac{\lambda^2}{ 4} \pd_x^3R_n \biggr).
\end{split}
\ee
Furthermore,
the free energy satisfies the puncture equation:
\be
\frac{\pd F}{\pd  t_{0}}
= \sum_{k=1}^\infty t_k\frac{\pd F}{\pd t_{k-1}} + \frac{t_0^2}{2\lambda^2},  \label{eqn:Puncture} \\
\ee
One can show that \eqref{eqn:KdV} and \eqref{eqn:Puncture} implies
the following equation \cite{DVV}:
\be \label{eqn:String}
u = \sum_{k=1}^\infty t_k  R_k + x.
\ee
This is called the {\em string equation} in Dijkagraaf-Verlinde -Verlinde \cite{DVV},
and is used by them to derive the Virasoro constraints.

\subsection{Kac-Schwarz characterization of $Z_{WK}$}
After the following change of coordinates:
\be \label{eqn:t-T}
t_n = (2n+1)!! T_{2n+1},
\ee
$Z_{WK}$ becomes a series in $T_1, T_3. \dots$,
and a tau-function of the KP hierarchy.
According to Kac-Schwarz \cite{Kac-Sch},
the element  $V \in \Gr_{(0)}$ corresponding to $Z_{WK}$ is given by a basis of the form
$\{z^{2n} a(z), z^{2n+1} b(z)\}_{n \geq 0}$,
where  $a(z)$ and $b(z)$ are given by the following series respectively:
\bea
&&  a(z) = \sum_{m=0}^\infty  \frac{(6m-1)!! }{6^{2m} (2m)!} z^{-3m}, \label{eqn:a} \\
&&  b(z) = - \sum_{m=0}^\infty  \frac{(6m-1)!!}{6^{2m} (2m)!}
\frac{6m+1}{6m-1} z^{-3m+1}. \label{eqn:b}
\eea
This space is characterized as follows:
\begin{align} \label{eqn:Kac-Schwarz}
z^2 V & \subset V, &  (\frac{1}{z}\pd_z - \frac{1}{2z^2} + z)V \subset V.
\end{align}

\subsection{Twistor data for topological 2D gravity}

By \eqref{eqn:Kac-Schwarz},
one gets the following twistor data for $Z_{WK}$:
\begin{align}
P & = L^2, & Q = L^{-1}M - \frac{1}{2} L^{-2} + L.
\end{align}

\subsection{The $\hat{S}$-operator for topological 2D gravity}

By the definition \eqref{def:S-Hat}, we get:
\be
\hat{S} = \sum_{n=1}^\infty  T_{n} P^{n/2}
+ \sum_{n =0}^\infty S_{n+1}(\bT; \hbar) P^{-n/2},
\ee
Then the noncommutative special deformation is given by:
\be
Q = \half \sum_{n=1}^\infty  n T_{n} P^{n/2-1}
-\half \sum_{n =0}^\infty S_{n+1}(\bT; \hbar) P^{-n/2-1}.
\ee
Since $Z_{WK}$ does not depend on $T_{2n}$ for $n \geq 1$,
we will set them to zero in the dispersionless version is given by:
\be
\cQ = \half \sum_{n=0}^\infty  (2n+1) T_{2n+1} \cP^{n-1/2}
+ \half \sum_{n=0}^\infty \frac{\pd F_0}{\pd T_{2n+1}} P^{-n-3/2}.
\ee
This is essentially \eqref{eqn:Special def} after a dilaton shift.

\subsection{Generalization to higher Weil-Petersson volumes}

Higher Weil-Petersson volumes
Consider also the following intersection numbers:
\be
\corr{\kappa_{b_1}\cdots \kappa_{b_k}
\tau_{a_1} \cdots \tau_{a_n}}_g: =
\int_{\Mbar_{g,n}} \kappa_{b_1} \cdots \kappa_{b_k}\psi_1^{a_1} \cdots \psi_n^{a_n},
\ee
where $\sum b_i + \sum a_j = 3g-3+n$.
When $a_1 = \cdots = a_n=0$,
these intersection numbers are called the higher Weil-Petersson volumes of $\Mbar_{g,n}$.
Consider the generating series of such intersection numbers:
\be
Z_{WP} =
\sum_g \corr{\exp (\sum_{k=1}^\infty s_k \kappa_k + \sum_{n=0}^\infty t_n \tau_n) }_g
\ee
It is known to be related to $Z_{WK}$ as follows
\cite{Kau-Man-Zag, Mul-Saf, Liu-Xu, Ber-Dub-Yan}:
\be
Z_{WP} = Z_{WK}(t_0, t_1, t_2-h_1(-\bs), \dots, t_{k+1} - h_k(-\bs), \dots),
\ee
where $h_(\bs)$ are polynomials in $s_1, s_2, \dots$ defined by
\be
\sum_{k=0}^\infty h_k(\bs) = \exp \biggl( \sum_{n=1}^\infty s_n x^n \biggr).
\ee
We define the special deformation of the spectral curve of $Z_{WP}$ by:
\be \label{eqn:Special def2}
x(\xi): = -\sum_{n \geq 0} \frac{t_n-h_j(-\bs)}{(2n-1)!!} \xi^{2n-1}
- \sum_{n \geq 0} (2n+1)!!\frac{\pd F_0}{\pd t_n}(\bt; \bs) \cdot \xi^{-2n-3},
\ee
where $h_{-1}(\bs) = 0$.
It follows that the spectral curve of $Z_{WP}$ can be obtained from the spectral curve
of $Z_{WK}$ through the special deformation of the spectral curve of $Z_{WK}$.

\subsection{Generalization to $r$-spin intersection numbers}

This is very straightforward.
The element $V$ in $\Gr^{(0)}$ for the generating series of $r$-spin intersection numbers \cite{Wit-R-Spin}
is characterized by \cite{Kac-Sch, Adl-van, Bal-Yan-Zho}:
\begin{align}
z^r V & \subset V, & (\frac{1}{z^{r-1}} \pd_Z - \frac{r-1}{2z^r} + z ) V & \subset V.
\end{align}
The twistor data is then:
\begin{align}
P & =L^r, &
Q = L^{1-r} M - \frac{r - 1}{2} \hbar L^{-r} + L.
\end{align}
So one should rewrite the $\hat{S}$-operator as:
\be
\hat{S}= \sum_{n=1}^\infty  T_{n} P^{n/r}
+ \sum_{n =0}^\infty S_{n+1}(\bT; \hbar) P^{-n/r},
\ee
and set $T_{rn} = 0$.
This will leads to deformation of the curve $y = \frac{1}{r} x^r$.
We will report the details in a separate paper.

\section{Conclusions}
\label{sec:CPS}

In this paper we have combined the twistor theory of KP hierarchy developed by Takasaki and Takebe \cite{Tak-Tak}
with the theory of string equations developed by Schwarz \cite{Sch}
into a theory for producing emergent mirror symmetry for topological matters coupled with 2D topological gravity.
For pure 2D gravity encoded in Witten-Kontsevich tau-function and for the generating
series of  higher Weil-Peterrson volumes,
one gets of quantum mechanics of the plane curve $y = \half x^2$ with its
deformation parameterized by big phase space of coupling constants of gravitation descendants;
for the series that encodes the $r$-spin intersection numbers, the curve is $y = \frac{1}{r} x^r$.

Our work can be modified and generalized in many different directions.
First of all,
one can relax the condition that either $P$ or $Q$ being a differential operator
of finite order.
Secondly, one can the case of $n$-component KP hierarchy.
Thirdly,
one can treat  a general integrable hierarchy by embedding it into an $n$-component KP hierarchy,
if it is possible.
To summarize,
the two diagrams in the Introduction can be generalized.
We will report these in subsequent publications.

Finally,
this work provides a way to geometrically engineer quantum mechanics
with some special actions
from topological matters coupled to topological 2D gravity.
One can expect that by taking other topological matters one can engineer more interesting  actions.
Of course 2d gravity is just a toy model,
a physical theory should be in 4D.
One can speculate about the plausibility of geometric engineering quantum mechanics
from 4D gravity coupled with matters in a similar fashion,
keeping in mind that twistor theory, a key ingredient in this work,
has its roots in 4D theory of instantons and gravitational instantons.
If this becomes successful,
then one gets a new way to relate gravity to quantum mechanics.

\vspace{.1in}

{\em Acknoledgements}.
This research is partially supported by NSFC grant 11171174.
The author thanks Professors Ference Balogh and Di Yang for collaborations on r-spin intersection numbers 
which are very helpful for this work.
He also thanks Professor Si Li for helpful discussions.

\end{document}